\documentclass{jfm}

\newcommand{\Aavg}[1]{\langle {#1} \rangle_{A^2}}

\usepackage{graphicx}
\usepackage{epstopdf,epsfig}
\usepackage{newtxtext}
\usepackage{newtxmath}
\usepackage{natbib}
\usepackage{xpatch}
\usepackage[colorlinks=true, urlcolor=blue, allcolors=blue]{hyperref}
\hypersetup{
    colorlinks = true,
    urlcolor   = blue,
    citecolor  = blue,
}

\usepackage{subfig}
\usepackage{relsize}
\usepackage{orcidlink}
\usepackage{tikz}
\usepackage{xcolor}
\usepackage{colortbl}
\usepackage{comment}
\usepackage{soul}
\usepackage[nameinlink,noabbrev]{cleveref}

\newcommand{\copyedit}[2][green]{{\colorlet{foo}{#1}\sethlcolor{foo}\hl{#2}}}
\newcommand{\rev}[1]{\textcolor{red}{#1}}
\renewcommand\copyedit[1]{#1}
\renewcommand\rev[1]{#1}

\newcommand{\RomanNumeralCaps}[1]
\linenumbers

\makeatletter
\xpatchcmd\NAT@citex
 {%
  \@citea\NAT@hyper@{%
    \NAT@nmfmt{\NAT@nm}%
    \hyper@natlinkbreak{\NAT@aysep\NAT@spacechar}{\@citeb\@extra@b@citeb}%
    \NAT@date
  }%
 }
 {%
  \@citea
  \NAT@nmfmt{\NAT@nm}%
  \NAT@aysep\NAT@spacechar
  \NAT@hyper@{\NAT@date}%
 }
 {}{}
\xpatchcmd\NAT@citex
 {%
  \@citea\NAT@hyper@{%
    \NAT@nmfmt{\NAT@nm}%
    \hyper@natlinkbreak{\NAT@spacechar\NAT@@open\if*#1*\else#1\NAT@spacechar\fi}%
    {\@citeb\@extra@b@citeb}%
    \NAT@date
  }%
 }
 {
  \@citea
    \NAT@nmfmt{\NAT@nm}%
    \NAT@spacechar\NAT@@open\if*#1*\else#1\NAT@spacechar\fi
    \NAT@hyper@{\NAT@date}%
 }
 {}{}
\makeatother

\title{Velocity gradient partitioning in turbulent flows}

\shorttitle{Velocity gradient partitioning in turbulent flows}
\shortauthor{R. Arun and T. Colonius}

\author{Rahul Arun\aff{1}\corresp{\email{\href{mailto:rarun@caltech.edu}{rarun@caltech.edu}}} \and Tim Colonius\aff{2}}

\affiliation{\aff{1}Graduate Aerospace Laboratories, California Institute of Technology, Pasadena, CA 91125, USA
\aff{2}Department of Mechanical and Civil Engineering, California Institute of Technology, Pasadena, CA 91125, USA}

\begin{document}
\maketitle

\vspace*{5ex}
\begin{abstract}
The velocity gradient tensor can be decomposed into normal straining, pure \copyedit{shearing} and rigid rotation tensors, each with distinct symmetry and normality properties. We partition the strength of \rev{turbulent velocity gradients} based on the relative contributions of these constituents in several canonical flows. These flows include forced isotropic turbulence, turbulent \copyedit{channels} and turbulent boundary layers. For forced isotropic turbulence, the partitioning is in excellent agreement with previous results. For wall-bounded turbulence, the partitioning collapses onto the isotropic partitioning far from the wall, where the mean shearing is relatively weak. By contrast, the near-wall partitioning is dominated by shearing. Between these two regimes, the partitioning collapses well at sufficiently high friction Reynolds numbers and its variations in the buffer layer and the log-law region can be reasonably \copyedit{modelled} as a function of the mean shearing strength. Altogether, our results highlight the \rev{expressivity and} broad applicability of the velocity gradient partitioning \rev{as advantages for} turbulence \copyedit{modelling}.
\end{abstract}




\section{Introduction}\label{sec:intro}

\rev{Identifying universal features of turbulent flows is a longstanding motif in turbulence \copyedit{modelling}. Kolmogorov's local isotropy and similarity hypotheses \citep{Kol1941} and their refinements \citep{Kol1962,Obo1962} suggest that small-scale flow statistics are approximately isotropic at sufficiently high Reynolds numbers, irrespective of the flow structure at large scales. Velocity gradients provide a useful testbed for investigating these hypotheses since they describe fundamental statistical and structural features of small-scale turbulence \citep{Men2011,Joh2024}. For example, \citet{Joh2017} found support for the local isotropy hypothesis sufficiently far from the wall in a turbulent channel using velocity gradient statistics related to vortex stretching. Aiming to inform turbulence \copyedit{modelling} efforts, the present study evaluates velocity gradient statistics in various wall-bounded flows relative to their isotropic values using a \copyedit{recently developed} normality-based analysis.}

Decomposing the velocity gradient tensor (VGT) based on its symmetry and normality properties distinguishes contributions from three distinct modes of deformation. These modes of deformation are normal straining, rigid \copyedit{rotation} and pure shearing. Originally, these modes were determined by applying a `triple decomposition' of the VGT in a `basic' reference frame where the effects of pure shearing can be extracted \rev{as a purely asymmetric tensor} \citep{Kol2007}. While identifying such a frame originally required solving challenging pointwise optimization problems, more computationally practical approaches based on the real Schur decomposition of the VGT have been developed recently \citep{Liu2018,Gao2018,Gao2019}. \rev{We adopt this more recent \copyedit{normality-based} triple decomposition, which identifies normal straining as symmetric and normal, rigid rotation as antisymmetric and normal, and pure shearing as strictly non-normal.} A complex Schur decomposition has also been used \citep{Key2018}, but its relationship to basic reference frames in physical space is less clear \citep{Kro2023}.

Partitioning the strength of the VGT based on its triple decomposition provides an expressive description of flow features. For example, the contribution of rigid rotation has been used to identify vortices \citep{Liu2018,Gao2018,Hal2021} and that of shearing has been found to leave a strong imprint on extreme velocity gradients associated with intermittency \citep{Das2020}. The interplay between shearing and rigid rotation has been used to characterize the transition and turbulent decay of colliding vortex rings mediated by the elliptic instability \citep{Aru2024}. \rev{The original triple decomposition of the VGT \citep{Kol2007} has also been used to \copyedit{analyse} the statistical and structural imprint of turbulent shearing \citep{Nag2020,Wat2020,Eno2023}. A theme of these works is the association of shearing with sheet-like vorticity and rigid rotation with tube-like vorticity.}

\citet{Das2020} showed that the velocity gradients in forced isotropic turbulence converge to a specific partitioning at high Taylor-scale Reynolds numbers. More recently, \citet{Aru2024} found that the decaying turbulent cloud produced by a vortex ring collision follows a similar partitioning. However, the spatial development in flows such as wakes, axisymmetric jets, and mixing layers has been associated with enhanced contributions of non-normal velocity gradients \citep{Bea2019}.

Using the \copyedit{normality-based} triple decomposition, we partition the strength of \rev{velocity gradients} in forced isotropic turbulence, turbulent \copyedit{channels} and turbulent boundary layers. The partitioning framework is presented in \textsection \ref{sec:math} and the turbulence datasets we \copyedit{analyse} are reported in \textsection \ref{sec:data}. We establish the isotropic partitioning in \textsection \ref{sec:eq} and thereafter discuss how the partitioning is modified for wall-bounded turbulence in \textsection \ref{sec:wall}, emphasizing the role of the mean shearing and the friction Reynolds number.

 \section{Partitioning framework}\label{sec:math}

The VGT, $\mathsfbi{A} = \boldsymbol{\nabla} \boldsymbol{u}$, can be expressed in its principal reference frame, denoted by \copyedit{$(\;\cdot\;)^*$}, as $\mathsfbi{A}^* = \mathsfbi{Q}\mathsfbi{A}\mathsfbi{Q}^{\rm T}$, where $\mathsfbi{Q}$ is unitary and \copyedit{$(\;\cdot\;)^{\rm T}$} represents the transpose. In this frame, the VGT is quasi-triangular and \rev{it can be decomposed as}
\begin{equation}\label{eq:VGT_triple}
    \mathsfbi{A}^* = \underbrace{\begin{bmatrix} \dot{\epsilon}_1^* & 0 & 0 \\ 0 & \dot{\epsilon}_2^* & 0 \\ 0 & 0 & \dot{\epsilon}_3^* \end{bmatrix}}_{\textstyle \mathsfbi{A}_\epsilon^*} + \underbrace{\begin{bmatrix} 0 & 0 & 0 \\ 0 & 0 & \dot{\varphi}_1^* \\ 0 & -\dot{\varphi}_1^* & 0 \end{bmatrix}}_{\textstyle \mathsfbi{A}_\varphi^*} + \underbrace{\begin{bmatrix} 0 & \dot{\gamma}_3^* & \dot{\gamma}_2^* \\ 0 & 0 & \dot{\gamma}_1^* \\ 0 & 0 & 0 \end{bmatrix}}_{\textstyle \mathsfbi{A}_\gamma^*},
\end{equation}
where $\mathsfbi{A}_\epsilon^*$, $\mathsfbi{A}_\varphi^*$ \copyedit{and} $\mathsfbi{A}_\gamma^*$ denote the normal straining, rigid \copyedit{rotation} and pure shearing tensors, respectively. These tensors can be determined and transformed to the original coordinate system using the ordered real Schur decomposition of $\mathsfbi{A}$ \citep{Kro2023}.

Correspondingly, the strength of the velocity gradients can be expressed as
\begin{equation}\label{eq:VGT_part}
    A^2 = {\rm tr}\left( \mathsfbi{A}^{\rm T} \mathsfbi{A}^{} \right) = \underbrace{{\rm tr}\left( \mathsfbi{A}_\epsilon^{\rm T} \mathsfbi{A}_\epsilon^{} \right)}_{\textstyle A_\epsilon^2} + \underbrace{{\rm tr}\left( \mathsfbi{A}_\varphi^{\rm T} \mathsfbi{A}_\varphi^{} \right)}_{\textstyle A_\varphi^2} + \underbrace{{\rm tr}\left( \mathsfbi{A}_\gamma^{\rm T} \mathsfbi{A}_\gamma^{} \right)}_{\textstyle A_\gamma^2} + \underbrace{2 {\rm tr}\left( \mathsfbi{A}_\varphi^{\rm T} \mathsfbi{A}_\gamma^{} \right)}_{\textstyle A_{\varphi\gamma}^2},
\end{equation}
where \copyedit{${\rm tr}(\;\cdot\;)$} represents the trace. The first three terms represent the strengths of the constituents in (\ref{eq:VGT_triple}) and the last term represents the interaction between shearing and rigid rotation. The velocity gradient partitioning is defined in terms of the relative contributions of these constituents to $A^2$. In normalized form, these contributions are bounded as $A_\epsilon^2 / A^2 \in [0,1]$, $A_\varphi^2 / A^2 \in [0,1]$, $A_\gamma^2 / A^2 \in [0,1]$ \copyedit{and} $A_{\varphi\gamma}^2 / A^2 \in [0,(\sqrt{2}+1)^{-1}]$ \citep{Das2020}.

Ensemble averaging can be used to characterize the statistical relevance of the contributions in this partitioning. We define the averaged partitioning as
\begin{equation}\label{eq:avg}
    \Aavg{A_\zeta^2}^{} = \langle A_\zeta^2 \rangle \big/ \langle A^2 \rangle, \quad \zeta \in \{\epsilon,\varphi,\gamma,\varphi\gamma\},
\end{equation}
where \copyedit{$\langle (\;\cdot\;) \rangle$} denotes averaging over homogeneous spatial directions and time. The present study focuses on this averaged partitioning, which can be used to characterize contributions to enstrophy and dissipation \citep{Das2020,Aru2024}.

\rev{The practical relevance of the partitioning is tied, in part, to its ability to inform \copyedit{modelling} efforts, including for non-canonical flows. Some \copyedit{modelling} paradigms, such as large-eddy simulations (LES) and Lagrangian approaches, operate on the basis of the total velocity. Other paradigms, such as the Reynolds-averaged Navier--Stokes (RANS) equations and input-output analyses, operate on the basis of the velocity fluctuations. In these contexts, it is therefore important to distinguish the partitioning of the total velocity gradients from the partitioning of the velocity gradient fluctuations, which we denote as $\Aavg{A_\zeta^2}^{}$ and $\Aavg{A_\zeta^2}'$, respectively. This distinction is particularly relevant since the normality-based triple decomposition does not generally commute with filtering or averaging operations. Physically, the modes of deformation for the total partitioning reflect what a fluid parcel would actually experience. By contrast, the modes for the fluctuation partitioning reflect what it would experience if advected only by the velocity fluctuations. We consider both the total partitioning and the fluctuation partitioning for the wall-bounded flows in the present study.}

 \section{Turbulence datasets}\label{sec:data}

\begin{table}
\begin{center}
\def~{\hphantom{0}}
\begin{tabular}{lclcrl}
Case   & Configuration    & Reynolds number                 & Grid size                   & $N_t\;\;\;$ & Reference       \\[0pt]
       &                  &                                 &                             &             &                 \\[-4pt]
FIT315 & Forced isotropic & $Re_\lambda \approx$ 315        & $(1024,1024,1024)$          & $67$        & \citet{Car2017} \\
FIT610 & Forced isotropic & $Re_\lambda \approx$ 610        & $(4096,4096,4096)$          & $1$         & \citet{Yeu2012} \\
Ch0186 & Channel          & $Re_\tau    \approx$ 186        & $(\quad32,\;\;129,\quad32)$ & $55\,925$   & \citet{Aru2023} \\
Ch1000 & Channel          & $Re_\tau    \approx$ 1000       & $(2048,\;\;512,1536)$       & $40$        & \citet{Gra2016} \\
BL0729 & Boundary layer   & $Re_\tau    \approx$ 292$-$729  & $(2049,\quad90,\;\;256)$    & $10\,000$   & \citet{Tow2023} \\
BL1024 & Boundary layer   & $Re_\tau    \approx$ 481$-$1024 & $(4097,\quad90,\;\;512)$    & $1500$      & \citet{Tow2023} \\
\end{tabular}
\caption{Turbulence datasets considered in the present analyses. Taylor-scale and friction Reynolds numbers are denoted by $Re_\lambda$ and $Re_\tau$, respectively, and $N_t$ denotes the number of snapshots. The grid sizes correspond to the streamwise ($x$), wall-normal ($y$), and spanwise ($z$) directions, respectively.}
\label{tab:data}
\end{center}
\end{table}

As summarized in \cref{tab:data}, we \copyedit{analyse} the partitioning in several well-validated turbulence datasets \rev{obtained from direct numerical simulations}. These datasets include forced isotropic turbulence (FIT315 and FIT610) and wall-bounded turbulence (Ch0186, Ch1000, \copyedit{BL0729} and BL1024) over a broad range of Reynolds numbers. \copyedit{Cases} FIT610 and Ch1000 are obtained from the Johns Hopkins Turbulence Databases \citep{Li2008} and Ch0186 represents a minimal flow unit for near-wall turbulence \citep{Jim1991}. The references in \cref{tab:data} provide further computational details and validation for each dataset.

We establish the isotropic partitioning using FIT610 and test its sensitivity to $Re_\lambda$ using FIT315. The snapshots for FIT315 are spaced roughly one integral time unit apart and $N_t$ is selected to produce a similar number of samples to FIT610.
We use Ch1000 to characterize the partitioning in a turbulent channel at a moderate $Re_\tau$. Its snapshots are spaced roughly 0.65 eddy turnover time units apart and they span roughly one flow-through time unit \citep{Gra2016}. \copyedit{Case} Ch0186 allows us to investigate how the wall-bounded partitioning changes when $Re_\tau$ is barely large enough to sustain turbulence. Its snapshots span roughly 160 eddy turnover time units.
We use BL0729 and BL1024 to further characterize the partitioning for wall-bounded turbulence subject to mild spatial development. Their broad ranges of $Re_\tau$ allow us to characterize how the partitioning evolves as the flows become increasingly turbulent. The snapshots for BL0729 and BL1024 span more than 20 and 7 eddy turnover time units, respectively \citep{Tow2023}.

When computing the VGT, we adopt the differentiation techniques employed in the original simulations where possible. We further require that all elements of the VGT be collocated prior to partitioning. For FIT315 and FIT610, we use a spectral method to compute all velocity gradients. For Ch0186, we use the second-order accurate staggered finite differences employed in the original simulation and subsequently shift staggered quantities to cell \copyedit{centres}. This shifting is performed by adjusting the phases of the Fourier modes in $x$ and $z$ and by averaging adjacent values in $y$ \citep{Aru2023}. For Ch1000, we use a spectral method in $x$ and $z$ and collocated finite differences with a stencil size of $N_s = 7$ in $y$. The original BL0729 and BL1024 simulations employed second-order accurate staggered finite differences; however, the published datasets are collocated and subsampled by a factor of two in $y$ and $z$ \citep{Tow2023}. As a result, we employ collocated finite differences with $N_s = 3$ in $x$ and $y$. Since the $z$ direction has periodic boundary conditions, we compute spanwise derivatives using a spectral method with the modified wavenumbers associated with the original staggered finite difference scheme.

\section{Partitioning in \copyedit{nearly isotropic} turbulence}\label{sec:eq}

\definecolor{dgray}{rgb}{0,0,0}
\definecolor{dred}{rgb}{1,0,0}
\definecolor{dblue}{rgb}{0,0,1}
\definecolor{dgreen}{rgb}{0,0.9,0}
\colorlet{lgray}{dgray!20}
\colorlet{lred}{dred!30}
\colorlet{lblue}{dblue!30}
\colorlet{lgreen}{dgreen!30}

\begin{table}
\setlength\tabcolsep{2.5pt}
\begin{center}
\def~{\hphantom{0}}
\begin{tabular}{lccccccccc}
Case & $\Delta_{iso}^{}|\Delta_{iso}'$ & \cellcolor{lgray} $\Aavg{A_\epsilon^2}^{}$ & \cellcolor{lred} $\Aavg{A_\varphi^2}^{}$ & \cellcolor{lblue} $\Aavg{A_\gamma^2}^{}$ & \cellcolor{lgreen} $\Aavg{A_{\varphi\gamma}^2}^{}$ & \cellcolor{lgray} $\Aavg{A_\epsilon^2}'$ & \cellcolor{lred} $\Aavg{A_\varphi^2}'$ & \cellcolor{lblue} $\Aavg{A_\gamma^2}'$ & \cellcolor{lgreen} $\Aavg{A_{\varphi\gamma}^2}'$ \\[3pt]
       &                   &                         &                        &                         &                          &                         &                        &                         &                          \\[-4pt]
FIT315 & $0.2\%|0.2\%$     & \cellcolor{lgray} 0.239 & \cellcolor{lred} 0.106 & \cellcolor{lblue} 0.521 & \cellcolor{lgreen} 0.134 & \cellcolor{lgray} 0.239 & \cellcolor{lred} 0.106 & \cellcolor{lblue} 0.521 & \cellcolor{lgreen} 0.134 \\
FIT610 & $-\;\;\;|\;\;-\;$ & \cellcolor{lgray} 0.240 & \cellcolor{lred} 0.106 & \cellcolor{lblue} 0.520 & \cellcolor{lgreen} 0.134 & \cellcolor{lgray} 0.240 & \cellcolor{lred} 0.106 & \cellcolor{lblue} 0.520 & \cellcolor{lgreen} 0.134 \\
Ch0186 & $5.0\%|5.0\%$     & \cellcolor{lgray} 0.250 & \cellcolor{lred} 0.090 & \cellcolor{lblue} 0.535 & \cellcolor{lgreen} 0.125 & \cellcolor{lgray} 0.250 & \cellcolor{lred} 0.090 & \cellcolor{lblue} 0.535 & \cellcolor{lgreen} 0.125 \\
Ch1000 & $0.4\%|0.4\%$     & \cellcolor{lgray} 0.242 & \cellcolor{lred} 0.105 & \cellcolor{lblue} 0.519 & \cellcolor{lgreen} 0.134 & \cellcolor{lgray} 0.242 & \cellcolor{lred} 0.105 & \cellcolor{lblue} 0.519 & \cellcolor{lgreen} 0.134 \\
BL0729 & $1.2\%|1.0\%$     & \cellcolor{lgray} 0.242 & \cellcolor{lred} 0.101 & \cellcolor{lblue} 0.524 & \cellcolor{lgreen} 0.133 & \cellcolor{lgray} 0.245 & \cellcolor{lred} 0.104 & \cellcolor{lblue} 0.517 & \cellcolor{lgreen} 0.134 \\
BL1024 & $0.8\%|0.6\%$     & \cellcolor{lgray} 0.239 & \cellcolor{lred} 0.102 & \cellcolor{lblue} 0.523 & \cellcolor{lgreen} 0.136 & \cellcolor{lgray} 0.242 & \cellcolor{lred} 0.104 & \cellcolor{lblue} 0.518 & \cellcolor{lgreen} 0.136 \\
\end{tabular}
\caption{Velocity gradient partitioning for each flow and the corresponding deviation metrics. The partitioning is reported at the channel \copyedit{centreline} for Ch0186 and Ch1000 and at \rev{$(Re_\tau,y^+) \approx (729,159)$ and $(1000,155)$} for BL0729 and BL1024, respectively. The column shadings reflect our partitioning \copyedit{colour} scheme.}
\label{tab:iso}
\end{center}
\end{table}

The isotropic velocity gradient partitioning characterizes the contributions of $\mathsfbi{A}_\epsilon$, $\mathsfbi{A}_\varphi$ \copyedit{and} $\mathsfbi{A}_\gamma$ in the idealized setting of forced isotropic turbulence. It has been established previously for \copyedit{$Re_\lambda \approx$ 1$-$588} \citep{Das2020} and is roughly invariant for $Re_\lambda \gtrsim 200$. Here, we use FIT610 to confirm this isotropic partitioning, which we denote by $\Aavg{A_\zeta^2}^{iso}$. We characterize deviations from this partitioning using the following metric\copyedit{:}
\begin{equation}
    \Delta_{iso}^{} = \sum\limits_{\zeta \in \{\epsilon,\varphi,\gamma,\varphi\gamma\}} \Big\lvert \langle A_\zeta^2 \rangle_{A^2}^{} - \langle A_\zeta^2 \rangle_{A^2}^{iso} \Big\rvert \Bigg/ \sum\limits_{\zeta \in \{\epsilon,\varphi,\gamma,\varphi\gamma\}} \Big\lvert \langle A_\zeta^2 \rangle_{A^2}^{iso} \Big\rvert,\label{eq:dev_iso}
\end{equation}
in which the denominator sums to unity. \rev{An analogous metric, $\Delta_{iso}'$, is defined for the fluctuation partitioning by replacing $\Aavg{A_\zeta^2}^{}$ with $\Aavg{A_\zeta^2}'$ in (\ref{eq:dev_iso}).} One advantage of these metrics is that they are not affected by further decomposing $\mathsfbi{A_\gamma}$ into its symmetric and antisymmetric parts. In the present study, they produce results similar to those produced by relative root-mean-square deviations, which are more commonplace.

\autoref{tab:iso} shows the partitioning alongside the deviation metrics for each dataset. Consistent with previous results \citep{Das2020}, the FIT315 partitioning is nearly identical to the FIT610 partitioning. For the wall-bounded flows, the partitioning is taken from regions where the mean shearing is relatively weak. Further, for the boundary layers, it is taken sufficiently far from the \copyedit{boundary-layer} thickness to mitigate the imprint of the exterior potential flow.

The partitioning in the selected regions of Ch1000, \copyedit{BL0729} and BL1024 is remarkably similar to the isotropic partitioning, with deviations of \rev{1.2\% or less. Since the mean flow has a minimal imprint on the velocity gradients in these regions, the total partitioning is quite similar to the fluctuation partitioning.} These results highlight that the isotropic partitioning is broadly applicable in appropriate regions of inhomogeneous turbulent flows.

 \section{Partitioning in wall-bounded turbulence}\label{sec:wall}

 \subsection{Effect of mean shearing}\label{sec:wall:meanShear}

\begin{figure}
    \centering
    \includegraphics[width=\textwidth]{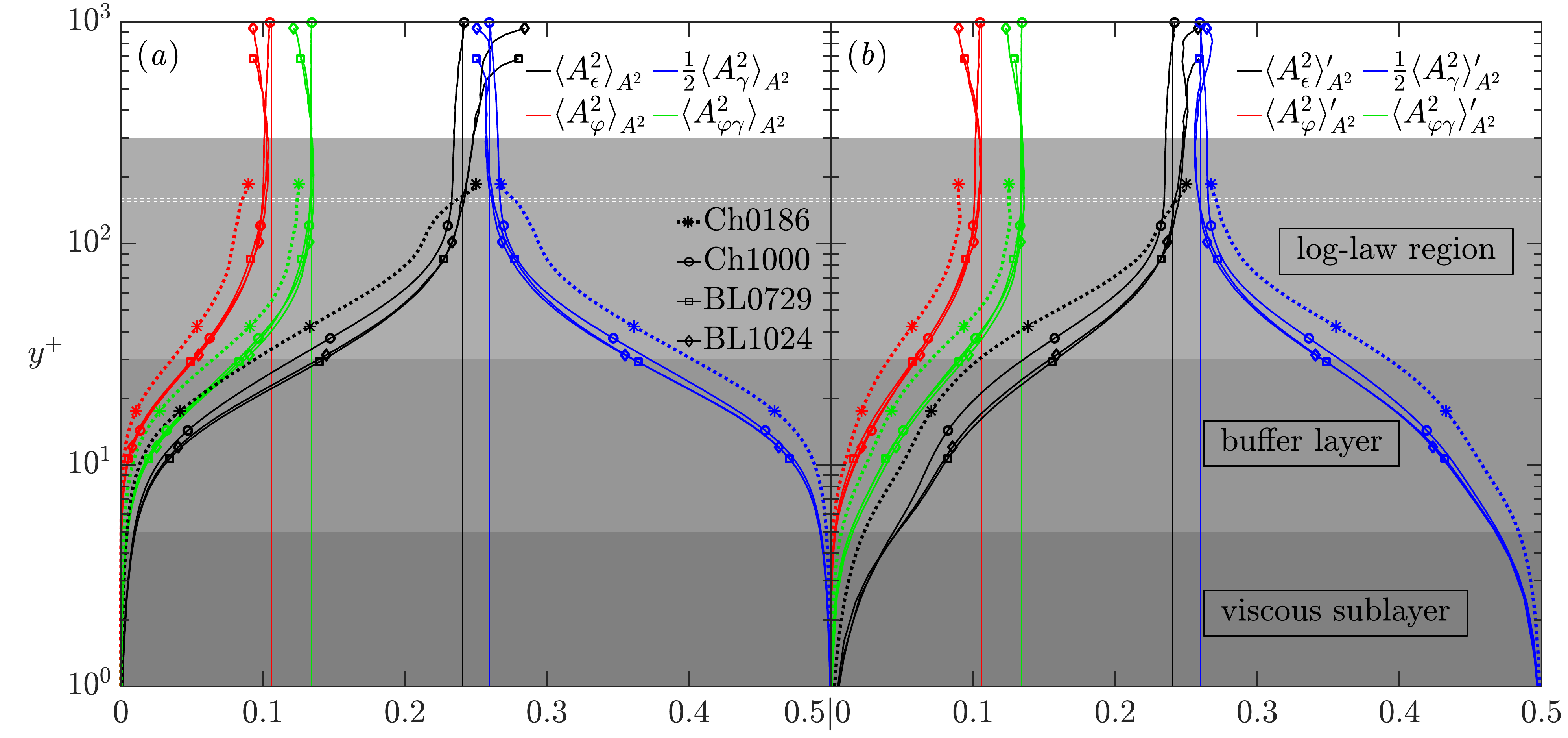}
    \includegraphics[width=\textwidth]{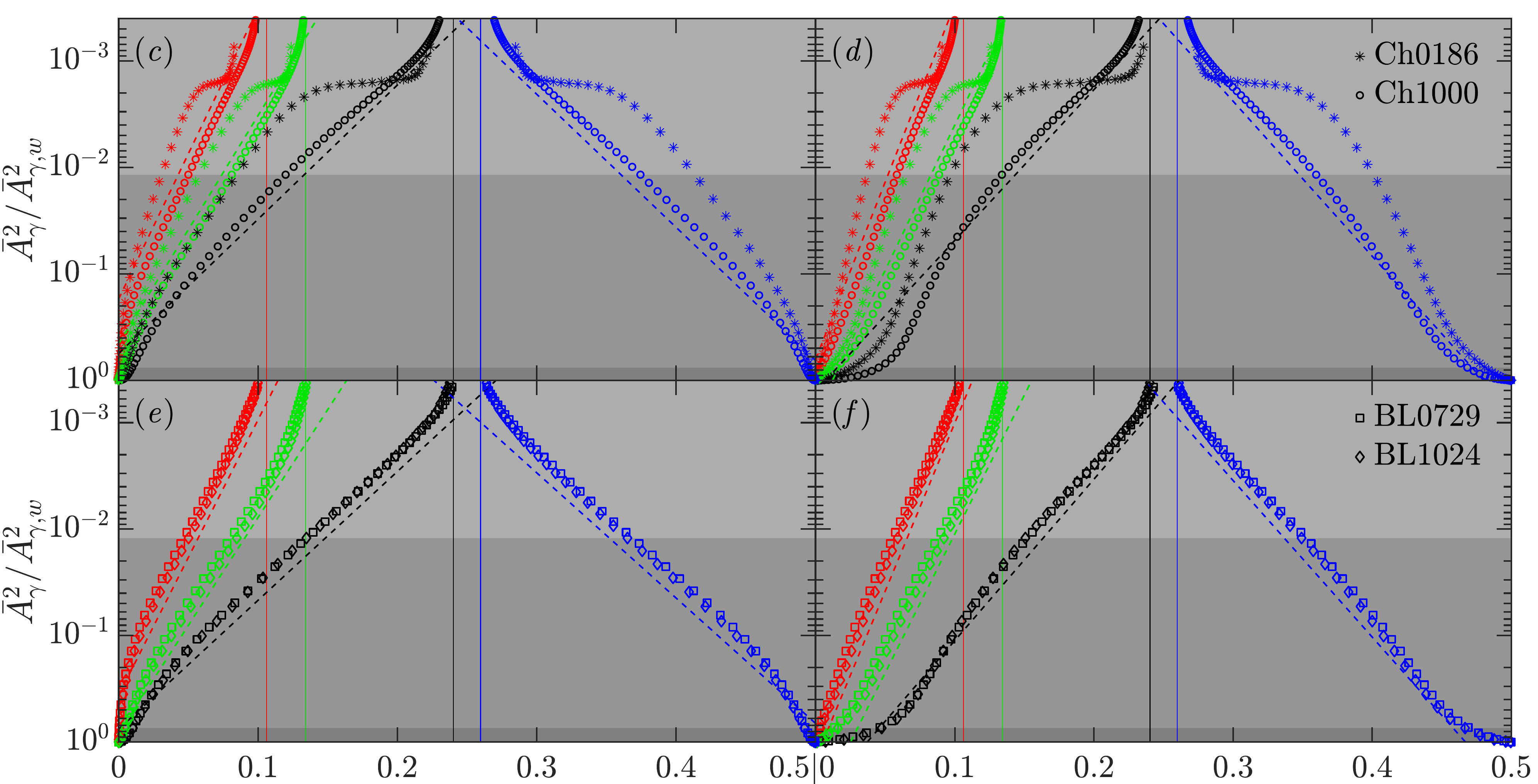}
    \caption{Total (\textit{a},\textit{c},\textit{e}) and fluctuation (\textit{b},\textit{d},\textit{f}) partitioning profiles for the channels and boundary layers in terms of wall-normal location in inner units (\textit{a},\textit{b}) and mean shearing strength (\textit{c}--\textit{f}). The vertical lines represent the isotropic values. The BL0729 and BL1024 profiles are shown for $Re_\tau \approx 729$ and $1000$, respectively, and the top boundary of the log-law region represents Ch1000. \rev{The dashed white lines in (\textit{a},\textit{b}) represent the locations of the partitioning values reported in \cref{tab:iso} for BL0729 and BL1024.} In (\textit{a},\textit{b}), the markers are used to distinguish between the profiles and, in (\textit{c}--\textit{f}), they represent actual data points. In (\textit{c}--\textit{f}), the mean shearing axis is reversed and the dashed lines represent comparable linear-log trends for each dataset, with the partitioning as the dependent variable.}
    \label{fig:profiles}
\end{figure}

The velocity gradient partitioning in wall-bounded turbulence is heavily influenced by the mean shearing imposed by the wall. \autoref{fig:profiles} shows the partitioning profiles as a function of both wall-normal location (in inner units) and mean shearing strength. Consistent with \cref{tab:iso}, the partitioning approaches the isotropic values far from the wall, starting \copyedit{near} the top of the log-law region. However, the \copyedit{boundary-layer} partitioning begins to diverge from these values near the edge of the boundary layer, reflecting the transition to a potential flow. Beyond the \copyedit{boundary-layer} thickness (not shown), this transition is associated with monotonic enhancement of normal straining and decay of the other constituents. As observed in \cref{tab:iso}, the Ch0186 partitioning does not converge as well onto the isotopic values since it is barely turbulent. The partitioning throughout this minimal channel is generally associated with enhanced contributions from shearing.

For all wall-normal profiles in \cref{fig:profiles}, the near-wall partitioning is dominated by shearing. This feature reflects the strong imprint of the mean shearing on the near-wall velocity gradients. \rev{The imprint of the mean shearing is also responsible for the enhanced contribution of shearing (at the expense of the other constituents) to the total partitioning relative to the fluctuation partitioning for $y^+ \lesssim 20$.} Between the near-wall regime and the \copyedit{nearly isotropic} regime far from the wall, the collapse of the partitioning profiles is particularly striking for BL0729, \copyedit{BL1024} and (to a lesser extent) Ch1000. The most significant differences between the channel and \copyedit{boundary-layer} profiles occur in the buffer layer. For the channels, the contributions of shearing and normal straining in this region are enhanced and reduced, respectively, relative to their contributions in the boundary layers. While beyond the scope of the present work, characterizing how the flow structures in the buffer layer reflect these differences \rev{(e.g. through the imprint of the exterior potential flow)} would be interesting future work.

The mean shearing strength profiles in \cref{fig:profiles} provide a complementary view to the wall-normal profiles. The mean shearing strength parameter is normalized as $\bar{A}_\gamma^2 / \bar{A}_{\gamma,w}^2$, where $\bar{A}_{\gamma,w}^2$ represents the (maximum) value at the wall. This parameter quantifies the effect of the wall in terms of velocity gradients and can be determined directly from the mean flow. Further, for the channels and boundary layers we consider, it can be well-approximated using \copyedit{the wall-normal gradient of the mean streamwise velocity}, $\partial \bar{u} / \partial y$. For these profiles, we \copyedit{focus primarily} on the buffer layer and log-law region since they capture the majority of the evolution from the near-wall regime to the \copyedit{nearly isotropic} regime. As observed for the wall-normal profiles, the partitioning in these regions collapses very well as a function of the mean shearing strength for BL0729, \copyedit{BL1024} and (to a lesser extent) Ch1000.

The partitioning in wall-bounded flows becomes similar to the isotropic partitioning when $\bar{A}_\gamma^2 / \bar{A}_{\gamma,w}^2 \lesssim 10^{-3}$. \rev{The dashed lines in \cref{fig:profiles} illustrate the similarity of the partitioning profiles to log-linear variations with the mean shearing strength for $10^{-3} \lesssim \bar{A}_\gamma^2 / \bar{A}_{\gamma,w}^2 \lesssim 10^{-1}$. The slopes of these variations for the total partitioning are slightly steeper than those for the fluctuation partitioning due to the enhanced contribution from the mean shearing for $\bar{A}_\gamma^2 / \bar{A}_{\gamma,w}^2 \gtrsim 10^{-1}$. The slopes for shearing and normal straining tend to be similar in magnitude and steeper than those of rigid rotation and shear-rotation interactions. This feature mirrors the relative contributions in the isotropic partitioning, for which $\Aavg{A_\epsilon^2}^{iso} \sim \tfrac{1}{2}\Aavg{A_\gamma^2}^{iso} \sim \Aavg{A_\varphi^2}^{iso} + \Aavg{A_{\varphi\gamma}^2}^{iso}$.}

More rigorous \copyedit{modelling} approaches may help enable predictions of the partitioning profiles in terms of mean flow variables. While we do not propose an explicit model for these profiles in the present study, our analysis suggests that \rev{both $y^+$ and} the mean shearing are appropriate for \copyedit{modelling} the partitioning and sufficiently high $Re_\tau$. \rev{Beyond mean flow variables, characterizing how the strength of the velocity gradient fluctuations relative to the mean shearing strength impacts these profiles may provide further insight.}

\rev{The pronounced, roughly monotonic variations of the normality-based partitioning profiles strikingly capture the development from the near-wall regime to the \copyedit{nearly isotropic} regime far from the wall. By contrast, as depicted and discussed in \autoref{sec:app:sec:A}, the variations for the symmetry-based partitioning are non-monotonic and do not exceed $\pm 0.02$ of their isotropic values in the regimes of interest. Supplemented by previous findings \citep{Aru2024}, these results highlight that the expressivity of the normality-based partitioning provides a key advantage over considering symmetry alone.}

 \subsection{Effect of friction Reynolds number}\label{sec:wall:Retau}

\begin{figure}
    \centering
    \includegraphics[width=\textwidth]{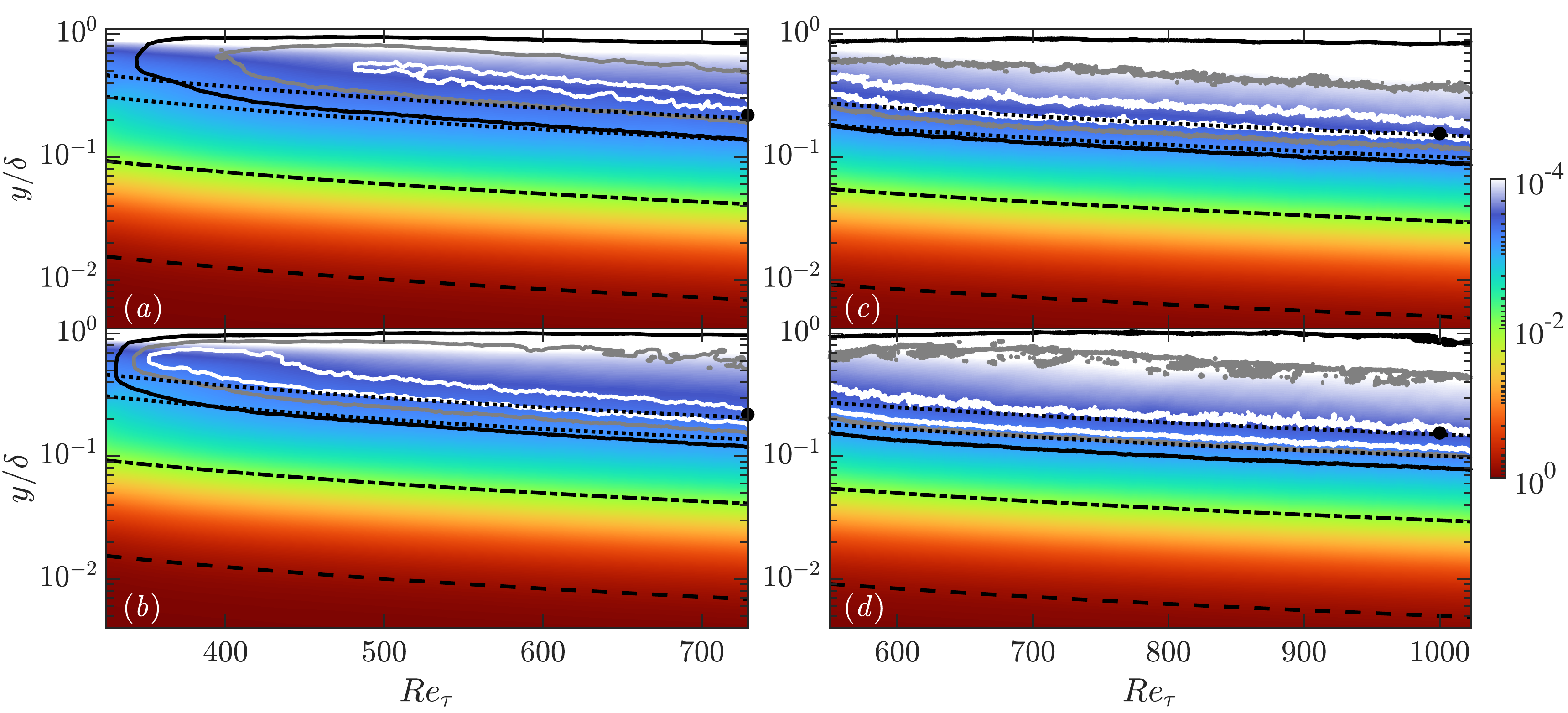}
    \caption{Streamwise development of BL0729 (\textit{a},\textit{b}) and BL1024 (\textit{c},\textit{d}) in terms of $Re_\tau$, where the \copyedit{colour} axis represents $\bar{A}_\gamma^2 / \bar{A}_{\gamma,w}^2$. \rev{The white, \copyedit{grey}, and black contours represent $\Delta_{iso}^{} = 1\%$, $\Delta_{iso}^{} = 2\%$ \copyedit{and} $\Delta_{iso}^{} = 5\%$, respectively, for the total partitioning (\textit{a},\textit{c}) and $\Delta_{iso}' = 1\%$, $\Delta_{iso}' = 2\%$ \copyedit{and} $\Delta_{iso}' = 5\%$, respectively, for the fluctuation partitioning (\textit{b},\textit{d}).} The dashed and dash-dotted black lines represent the top of the viscous sublayer $(y^+ = 5)$ and the top of the buffer layer $(y^+ = 30)$, respectively, and the dotted black lines represent \rev{$y^+ = 100$ and $y^+ = 150$}. The black circles represent the locations of the partitioning values reported in \cref{tab:iso} and $\delta$ represents the \copyedit{boundary-layer} thickness.}
    \label{fig:Retau}
\end{figure}

Beyond wall-normal variations, the streamwise spatial development in the boundary layers is associated with increasing $Re_\tau$. \autoref{fig:Retau} shows how the region where the isotropic partitioning is applicable evolves as a function of $Re_\tau$. For BL0729, this region grows appreciably in size as $Re_\tau$ increases. The same is true to a lesser extent for BL1024, where the turbulence is more well-developed throughout the domain. The $\Delta_{iso}^{}$ \rev{and $\Delta_{iso}'$} contours identify regions where the isotropic partitioning is most applicable. Throughout BL1024 and at the downstream end of BL0729, it is most applicable around $y^+ \approx 170$ and $y^+ \approx 140$ for the total and fluctuating velocity gradients, respectively. \rev{The domain of applicability of the isotropic partitioning is slightly larger for the velocity fluctuations than for the total velocity.}

\rev{Except at the upstream end of BL0729, $\Delta_{iso}^{}$ and $\Delta_{iso}'$ are less than roughly 5\% for $y^+ \gtrsim 100$. Alongside \cref{fig:profiles}, these results complement the findings of \citet{Joh2017}, which suggest that velocity gradient statistics associated with vortex stretching approach their isotropic values for $y^+ \gtrsim 100$. Therefore, our results further support their conclusion that this collapse provides support for the local isotropy hypothesis sufficiently far from the wall.}

The inner unit scaling of this collapse contrasts with the \copyedit{boundary-layer} thickness, which scales in outer units and captures the divergence from the isotropic partitioning near the \copyedit{free stream}. Our results therefore suggest that, while inner unit scalings can be used to determine when the partitioning approaches the isotropic values, outer unit scalings may more appropriately capture the divergence to a potential flow in the boundary layers. They further suggest that $Re_\tau \gtrsim 700$ is a reasonable regime in which to expect a collapsed partitioning. This estimate is consistent with the collapse of the partitioning profiles for BL0729, \copyedit{BL1024} and (to a lesser extent) Ch1000 as well as the lack of collapse for Ch0186 in \cref{fig:profiles}.

\section{Concluding remarks}\label{sec:conc}

We have \copyedit{analysed} the normality-based partitioning of \rev{velocity gradients} in several canonical turbulent flows. The partitioning we compute for forced isotropic turbulence agrees well with previous results \citep{Das2020}. Moreover, we show that the isotropic partitioning also applies to velocity gradients near and beyond the top of the log-law region in wall-bounded flows over a broad range of $Re_\tau$. \rev{The broad applicability of the isotropic partitioning for $y^+ \gtrsim 100$ complements previous results \citep{Joh2017}, thereby providing further support for the local isotropy hypothesis for well-developed turbulence far from solid boundaries.} Our results suggest that $Re_\tau \gtrsim 700$ is sufficiently high to expect the partitioning profiles to collapse as they transition from the shearing-dominated near-wall regime to the \copyedit{nearly isotropic} regime. Further, the mean shearing provides a reasonable mean flow parameter for \copyedit{modelling} their variations in the buffer layer and the log-law region. \rev{Altogether, our results highlight expressivity as a key advantage of the normality-based partitioning over symmetry-based approaches.}

Moving forward, \copyedit{analysing} the partitioning profiles at higher $Re_\tau$ would help further characterize their sensitivity and collapse for wall-bounded flows. Developing more rigorous models for the partitioning in terms of mean flow variables would also be useful, especially models that do not depend strongly on the flow configuration. \rev{The partitioning may aid turbulence \copyedit{modelling} efforts in RANS, LES, \copyedit{Lagrangian} or input-output settings, e.g. by directly informing closure models or by providing an evaluation metric for models of interest.} Finally, connecting the statistical features we report to the turbulence structures that produce them would provide an enhanced view of the roles of the partitioning constituents.

\vspace*{3.5ex}

\backsection[Acknowledgements]{The authors gratefully acknowledge the reviewers for helpful feedback, A. Nekkanti and M. Wadas for discussions and H.J. Bae for providing the Ch0186 dataset.}

\backsection[Funding]{R.A. was supported by the Department of Defense (DoD) through the National Defense Science \& Engineering Graduate (NDSEG) Fellowship Program.}

\backsection[Declaration of interests]{The authors report no conflict of interest.}

\backsection[Data availability statement]{Sample code for computing the velocity gradient partitioning is available at \href{https://doi.org/10.22002/17h15-gr910}{https://doi.org/10.22002/17h15-gr910}.}

\backsection[Author ORCIDs]{

\noindent \orcidlink{0000-0002-5942-169X} Rahul Arun \href{https://orcid.org/0000-0002-5942-169X}{https://orcid.org/0000-0002-5942-169X}

\noindent \orcidlink{0000-0003-0326-3909} Tim Colonius \href{https://orcid.org/0000-0003-0326-3909}{https://orcid.org/0000-0003-0326-3909}

}


\appendix

\section{Symmetry-based partitioning profiles}\label{sec:app:sec:A}

\begin{figure}
    \centering
    \includegraphics[width=\textwidth]{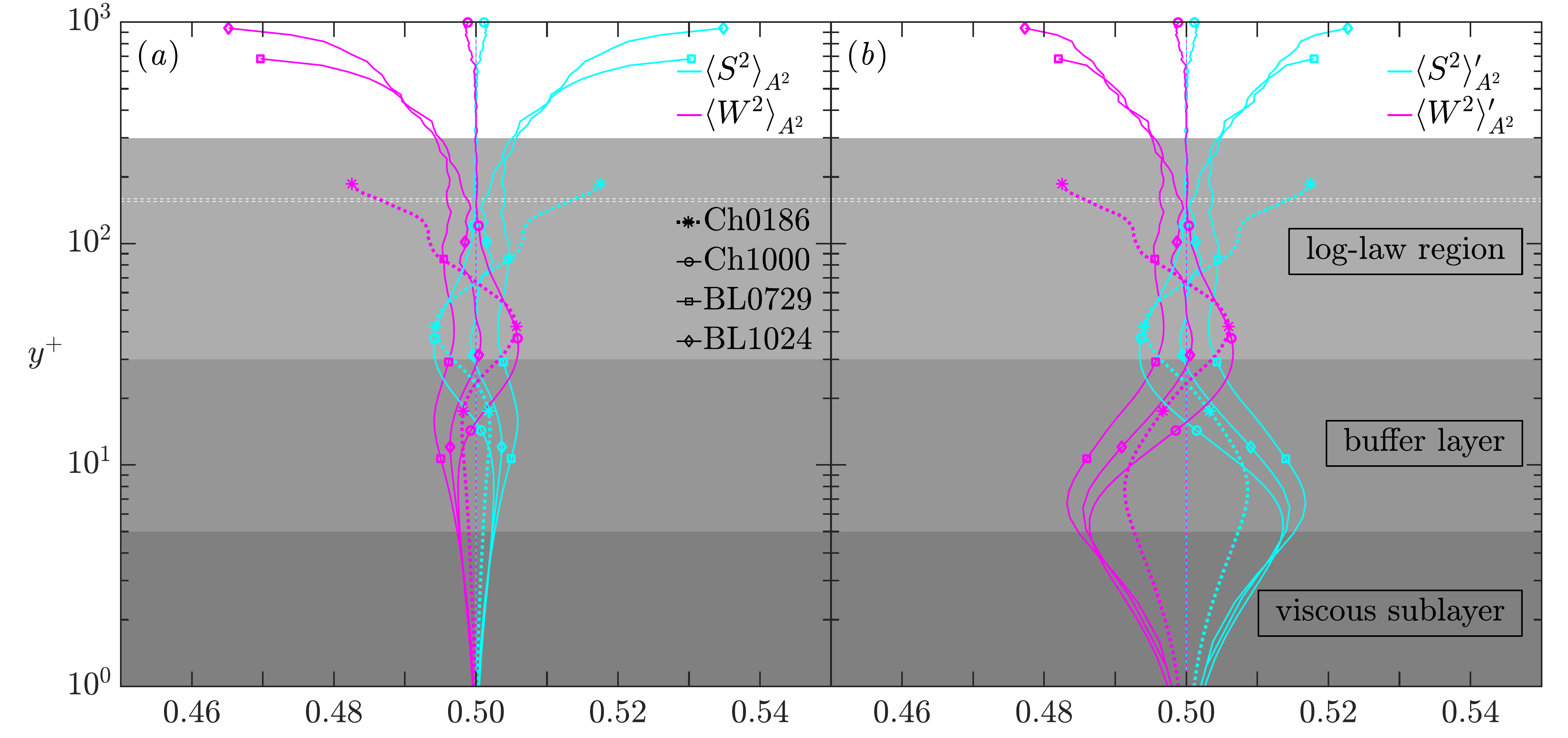}
    \caption{Symmetry-based total (\textit{a}) and fluctuation (\textit{b}) partitioning profiles for the channels and boundary layers in terms of wall-normal location in inner units. The plots are in the same style as those in \cref{fig:profiles}.}
    \label{fig:profiles_sym}
\end{figure}

\rev{The standard symmetry-based decomposition of the VGT identifies the strain-rate tensor as $\mathsfbi{S} = \tfrac{1}{2}(\mathsfbi{A} + \mathsfbi{A}^{\rm T})$ and the vorticity tensor as $\mathsfbi{W} = \tfrac{1}{2}(\mathsfbi{A} - \mathsfbi{A}^{\rm T})$. These tensors can be expressed in terms of the normality-based triple decomposition of the VGT as $\mathsfbi{S} = \mathsfbi{A}_\epsilon^{} + \mathsfbi{S}_\gamma^{}$ and $\mathsfbi{W} = \mathsfbi{A}_\varphi^{} + \mathsfbi{W}_\gamma^{}$, where $\mathsfbi{S}_\gamma^{} = \tfrac{1}{2}(\mathsfbi{A}_\gamma^{} + \mathsfbi{A}_\gamma^{\rm T})$ and $\mathsfbi{W}_\gamma^{} = \tfrac{1}{2}(\mathsfbi{A}_\gamma^{} - \mathsfbi{A}_\gamma^{\rm T})$.}

\rev{Using the definition in (\ref{eq:avg}), the symmetry-based partitioning characterizes the relative contributions of the strain-rate and vorticity tensors to the velocity gradient strength. In isotropic turbulence, these contributions are equipartitioned as $\Aavg{S^2}^{} = \Aavg{W^2}^{} = 0.500$. \autoref{fig:profiles_sym} shows the symmetry-based partitioning profiles for the wall-bounded flows we consider. Except near the potential flow regime of the boundary layers, these profiles do not deviate more than $\pm 0.02$ from the isotropic values for both the total and fluctuating velocities. Further, unlike the normality-based partitioning profiles in \cref{fig:profiles}, the symmetry-based profiles do not vary monotonically with $y^+$. These results show that the normality-based partitioning is significantly more expressive of the spatial variations in turbulence characteristics from the near-wall regime to the \copyedit{nearly isotropic} regime. This finding complements the results of \citet{Aru2024}, which show that the normality-based partitioning is more expressive than the symmetry-based partitioning in capturing the temporal evolution of a vortex ring collision, including its transition and turbulent decay. Together, these results highlight the superior expressivity of the normality-based partitioning.}


\vspace*{1.0ex}
\bibliographystyle{jfm}
\bibliography{references}

\end{document}